\documentclass[aps,prd,twocolumn,superscriptaddress,showpacs,nofootinbib]{revtex4-1}

\usepackage{multirow}
\usepackage{amsmath}
\usepackage{graphicx}
\usepackage{xcolor}
\usepackage{slashed}
\usepackage{hyperref}

\newcommand{\tr}{\text{tr}\,}
\newcommand{\Tr}{\text{Tr}\,}

\begin{document}


\title{Baryonic Matter Onset in Two-Color QCD with Heavy Quarks}


\newcommand{\affDARMSTADT}{Institut f\"ur Kernphysik, Technische Universit\"at Darmstadt, 64289 Darmstadt, Germany}
\newcommand{\affGIESSEN}{Institut f\"ur Theoretische Physik, Justus-Liebig-Universit\"at, 35392 Gie{\ss}en, Germany}


\author{Philipp Scior}
\email[]{scior@theorie.ikp.physik.tu-darmstadt.de}
\affiliation{\affDARMSTADT}

\author{Lorenz von Smekal}
\affiliation{\affDARMSTADT}
\affiliation{\affGIESSEN}


\date{\today}

\begin{abstract}
  We study the cold and dense regime in the phase diagram of two-color QCD with heavy quarks within a three-dimensional effective theory for Polyakov loops. This theory is derived from two-color QCD in a combined strong-coupling and hopping expansion. In particular, we study the onset of diquark density as the finite-density transition of the bosonic baryons in the two-color world. In contrast to previous studies of heavy dense QCD, our zero-temperature extrapolations are consistent with a continuous transition without binding energy. They thus provide  evidence that the effective theory for heavy quarks is capable of describing the characteristic differences between diquark condensation in two-color QCD and the liquid-gas transition of nuclear matter in QCD.
\end{abstract}


\maketitle

\section{Introduction}
	Tremendeous experimental and theoretical efforts have been devoted to the exploration of the QCD phase diagram over many years \cite{BraunMunzinger:2008tz}.
Lattice QCD studies were able to clarify the nature of the deconfinement phase transition for vanishing net-baryon density and calculate the equation of state to high accuracy \cite{Borsanyi2014,Bazavov:2014pvz}. A good description of the lattice data in this regime is nowadays being obtaind in Polyakov-loop extended chiral models \cite{Herbst:2013ufa}. Based on functional continuum methods for QCD such as the Functional Renormalization Group \cite{Mitter:2014wpa,Braun:2014ata} or Dyson-Schwinger equations \cite{Fischer2014} these studies can then be extended to finite baryon chemical potential \cite{Pawlowski:2014aha}. For sufficiently high temperatures one can furthermore use resummed perturbation theory to include baryon chemical potentials beyond Taylor-series expansions \cite{Mogliacci:2013mca}. 

The situation is different, however, in the low temperature and high baryon-density region of the QCD phase diagram as relevant for the properties of the inner cores of neutron stars. The fermion-sign problem impedes investigations of this region by standard Monte-Carlo simulations \cite{Forcrand2009}.  Methods to circumvent this problem are under active development: Complex Langevin dynamics has been shown to work for finite-density QCD with dynamical quarks at least on small lattices \cite{Sexty:2013ica}. Monte-Carlo on a Lefshetz thimble has been successfully tested for complex scalar fields with chemical potential \cite{Cristoforetti:2014gsa}. Dual lattice representations can help to overcome the sign problem for scalar fields and Abelian lattice models with worm algorithms \cite{Langfeld:2013kno,Mercado:2013ola,Bruckmann:2015sua}.
In a similar spirit, graph representations in terms of hadron worldlines can be used for lattice QCD with unrooted staggered quarks at finite density in combination with strong-coupling expansion techniques \cite{deForcrand:2009dh,deForcrand:2014tha}.

To get access to the continuum limit, one can combine strong-coupling and hopping expansions in order to derive three-dimensional effective Polyakov loop theories for heavy quarks \cite{Langelage2011,Fromm2012, Greensite2013}.
It has been shown, for example, that such effective theories can reproduce the critical couplings for deconfinement in the pure  $SU(2)$ and $SU(3)$ gauge theories, and the critical quark masses limiting the region of the first-order deconfinement transition in $SU(3)$ with heavy quarks, within less than 10\% \cite{Langelage2011,Fromm2012}. Studies of the cold and dense regime of the QCD phase diagram have produced first results on the nuclear-matter transition for QCD with heavy quarks \cite{Fromm:2012eb,Langelage}. In particular, the effective theory yields a binding energy per nucleon and a nuclear saturation density in the right ballpark relative to the heavy nucleon mass, with evidence of a first-order liquid-gas transition at very low temperatures which are currently still
beyond the range of applicability of the expansion, however.

 More direct evidence of a first-order transition analogous to that of nuclear matter has so far only been seen in $G_2$-QCD at finite density, a QCD-like theory with fermionic baryons but without sign problem which can therefore be simulated with standard lattice techniques \cite{Maas2012,Wellegehausen2014}. 
Another QCD replacement without sign problem is two-color QCD with the gauge group $SU(2)$. While it does not have fermionic baryons, it is much cheaper to simulate than $G_2$-QCD with 7 colors and 14 gluons corresponding to the fundamental representations of the smallest exceptional Lie group $G_2$ of rank 2 which contains the $SU(3)$ gauge group of QCD as a subgroup. 

The cold and dense regime in the phase diagram of two-color QCD with diquarks as bosonic baryons does not resemble the properties of nuclear matter but rather those of QCD at finite isospin density with charged pion condensation \cite{Kamikado:2012bt}. Detailed lattice results for the phase diagram of two-color QCD are available \cite{Hands2006,Cotter2012,Boz2013} and its qualitative features are captured by effective chiral models rather well \cite{Strodthoff:2013cua}. They reflect Bose-Einstein condensation of the baryonic diquarks in a second-order phase transition which occurs at zero temperature when the baryon chemical potential $\mu_B$ reaches the diquark mass, and the analog of a BEC-BCS crossover for larger values of $\mu_B$ as chiral symmerty gets gradually restored inside the diquark superfluid.

In this paper we apply the strong-coupling and hopping expansion techniques of Refs.~\cite{Fromm:2012eb,Langelage} to two-color QCD, in order to test to what extend these relatively well established features of the cold and dense regime of the two-color phase diagram are reflected in the effective theory for heavy quarks. Our results from the next-to-next-to-leading order in this expansion scheme  indicate that the corresponding three-dimensional effective Polyakov loop theory is capable of describing characteristic differences between two-color and real QCD at finite $\mu_B$. One can furthermore directly compare our results to lattice simulations of the full theory in two-color QCD at finite $\mu_B$ to assess where the effective theory eventually breaks down when going to smaller quark masses in the future.

        \section{Theoretical Background}
        \label{Background}
	\subsection{Effective Polyakov Loop Theory}{\label{ssub:effective_polyakov_loop_theory}}
We start with the derivation of the effective action. We will only briefly summarize the steps leading to the effective action. As most of the steps are very similar, we point the reader to a very detailed derivation for the the case of $SU(3)$ given in \cite{Langelage}. We start from 3+1 dimensional two-color QCD with $N_f$ flavors of Wilson fermions. After integration over the fermion field, the partition function reads
\begin{equation}
 	Z=\int [dU_ \mu] \; \exp[-S_\text{g}] \prod_{f=1}^{N_f} \det D_f \; ,
 \end{equation} 
with the fermion determinant $\det D_f$ and the Wilson gauge action
\begin{equation}
	S_\text{g}=\frac{\beta}{2 N_c}\sum_p (\tr U_p+ \tr U_p^\dagger)  \; .
\end{equation}
The effective action is then defined by integration over the spatial links
\begin{align}
Z&=\int [dU_0] \; \exp[-S_\text{eff}] \; , \notag \\ 
\exp[-S_\text{eff}]&= \int [dU_ i] \; \exp[-S_\text{g}] \prod_{f=1}^{N_f} \det D_f \; . \label{eff_action}
\end{align}
We can seperate the effective action
\begin{equation}
	S_\text{eff}(W)= S_1(W) + S_2(W) \; ,
\end{equation}
where $S_1$ contains all contributions from the Yang-Mills action potentially modified by contributions from non-winding fermion loops, while $S_2$ contains the winding fermion loops with gauge corrections. Both $S_1$ and $S_2$ now only depend on temporal Wilson lines or their traces, the Polyakov loops
\begin{equation}
L_{\vec x}= \tr W_{\vec x} =\tr \prod_{t=0}^{N_t-1} U_0(\vec x,t) \; .
\end{equation}
The easiest way to determine the contributions of $S_1$ is by using character expansion of $\exp[-S_\text{g}]$. In leading order this results in a nearest-neighbor interaction between Polyakov loops
\begin{align}
	S_1^{(\text{LO})}&= \lambda(\beta, N_t) \sum_{\langle \vec x \vec y\rangle} L_{\vec x} L_{\vec y} \; , \label{gauge_action} \\
	\lambda(u(\beta), N_t)&=u^{N_t}\exp\left[N_{t}\left(4u^4-4u^6+\mathcal{O}(u^8)\right)\right] \notag \; ,
\end{align}
with the fundamental coefficient of the character expansion $u(\beta)= I_2(\beta)/I_1(\beta) = 
\frac{\beta}{4}( 1 -\frac{ \beta^2}{24}+ \mathcal{O}(\beta^4))$.

The contributions of the fermion determinant to the effective action $S_2$ are determined by hopping expanding the fermion determinant. Here we again follow the example of \cite{Langelage} and split the determinant in temporal and spatial hops, $T=T^+ + T^-$ and $S=S^+ + S^-$ in forward and backward direction, 
\begin{align}
\det[D]&=\det[1-T^+-T^--S^+-S^-]  \notag \\
&=\det[1-T]\det[1-(1-T)^{-1}(S^++S^-)]  \notag \\
&=\det[D_\text{stat}]\det[D_\text{kin}] \; . \label{determinant}
\end{align} 
The part containing only temporal hops, called the static part of the determinant, is easy to evaluate since we do not have to integrate over spatial links. By evaluating the spin and space determinant we get
\begin{equation}
 	\det[D_\text{stat}]=\prod_{\vec x} (1+hL_{\vec x} +h^2)^2(1+\bar h L_{\vec x}  + \bar h^2)^2 \; ,
 \end{equation} 
 with
 \begin{equation}
 	h=(2 \kappa e^{a \mu})^{N_t}=e^{\frac{\mu-m_q}{T}} \hspace{0.2cm} \text{and} \hspace{0.2cm} \bar h = h(-\mu) \; . \label{ferm_coupling}
 \end{equation}
In the strong-coupling limit $\beta = 0$ the connection between the hopping parameter $\kappa$ and the constituent quark mass is then given by $am_q=-\ln(2 \kappa)$ \cite{Green1984}. To evaluate the kinetic quark determinant we further split the determinant in into parts describing quarks moving in positive $P_i$ and negative $M_i$ spatial directions $i$, with $P=\sum_i P_i = (1-T)^{-1} S^+ $ and $M=\sum_i M_i = (1-T)^{-1} S^-$, 
 \begin{align}
  \det[D_\text{kin}]&=\det[1-(1-T)^{-1}(S^+-S^-)]  \notag \\
  &=\det[1-P-M] \notag \\
 &=\exp[\Tr \log(1-P-M)] \; . \label{determinant2}
 \end{align}
 Since the trace in (\ref{determinant2}) is also a trace in coordinate space, only closed loops contribute for which we need equal numbers of $P$'s and $M$'s.  
 Up to order $\mathcal{O}(\kappa^4)$ this yields the following terms
 \begin{align}
   \det[D_{\text{kin}}]=& \notag \\
 &\hskip -1.6cm  \exp \left[ -\Tr PM -\Tr PPMM - 
  \frac{1}{2} \Tr PMPM +\mathcal{O}(\kappa^6) \right] \notag \\
   =&\,  1 -\Tr PM -\Tr PPMM - \frac{1}{2} \Tr PMPM \notag \\
   &  + \frac{1}{2} (\Tr PM)^2 +\mathcal{O}(\kappa^6) \, .  \label{det_kin}
 \end{align}
 When we evaluate the terms in (\ref{det_kin}) we recover all the terms that are there in the case of $SU(3)$. However there are two additional terms that appear for $SU(2)$ in the expressions
 \begin{align}
   -\frac{1}{2} \int \, dU \sum_{i} \tr P_i M_i P_i M_i \; , \;\;  \mbox{and}
   \notag \\
\frac{1}{2} \int \, dU \sum_{i} \tr P_i M_i \; \tr P_i M_i \; , 
\end{align}
because of an additional non-vanishing group integral
\begin{equation}
\int \; dU \, U_{ij} U_{kl}= \epsilon_{ik} \epsilon_{jl} \; . \label{groupintegral} 
\end{equation}
We identify those terms with possible diquark contributions in the two-color theory. With (\ref{groupintegral}) the contributions from the diquark terms become:
\begin{align}
-\frac{1}{2} \int \, dU \sum_{i}& \tr P_i M_i P_i M_i = \notag \\ &-16 \kappa^4 \sum_{x,y}\det[B_{x,y}] \det[B_{x+i,y+i}] \; , \notag \\
 \frac{1}{2} \int \, dU \sum_{i}& \tr P_i M_i \tr P_i M_i= \label{diquarkterms}\\ &32\kappa^4 \sum_{x,y}\det[B_{x,y}] \det[B_{x+i,y+i}] \; , \notag 
\end{align}
where $B_{x,y}$ are parts of the static quark propagator,
\begin{align}
D_\text{stat}^{-1}& = A + \gamma_0 B \; , \quad B_{x,y} = B^+_{x,y} + B^-_{x,y} \; , \\
B^+_{x,y}&=-\frac{1}{2} \frac{h W}{1+h W} \delta_{xy} \notag \\
  &+\frac{1}{2} z^{t_y-t_x} \frac{W(t_x,t_y)}{1+h W} \left[\theta(t_y-t_x) - h \theta(t_x-t_y) \right] \delta_{\vec x \vec y} \; , \notag \\
B^-_{x,y}&=- \frac{1}{2} \frac{\bar h W^\dagger}{1+ \bar h W^\dagger} \notag \\
&+\frac{1}{2} \bar z^{t_x-t_y} \frac{W^\dagger(t_x,t_y)}{1+\bar h W^\dagger} \left[\theta(t_x-t_y) - \bar h \theta(t_y-t_x) \right] \delta_{\vec x \vec y} \; ,  \notag 
\end{align}
and are defined in the same way as in \cite{Langelage} with $z= 2\kappa\, e^{a\mu}$ and $\bar z= 2\kappa\, e^{-a\mu}$.
Eqs.~(\ref{diquarkterms}) then lead to an overall di\-quark contribution of the form 
\begin{align}
-S_{\text{diquark}}&= 16 \kappa^4 \sum_{x,y}\det[B_{x,y}] \det[B_{x+i,y+i}]  \\
&= \kappa^4 N_t^2 \sum_{\vec x,i} h^4 \det \frac{1}{(1+h W_{\vec x})(1+h W_{\vec x+i})}  \notag \\
&= \kappa^4 N_t^2 \sum_{\vec x,i} \frac{h^4}{(1+h L_{\vec x}+h^2)(1+h L_{\vec x+i}+h^2)} \; . \notag
\end{align}

\subsubsection{Strong Coupling Limit and Gauge Corrections}{\label{ssub:strong_coupling}}
When leaving the strong-coupling limit $\beta=0$ one might think that we should include Polyakov-loop interactions of the form (\ref{gauge_action}). However, it turns out that the strong-coupling description is still valid even at $\beta=2.5$. This is due to the fact that the effective coupling for the gauge sector $\lambda$ is negligible in the temperature regime we are interested in ($T\leq 10$ MeV). For the temperature to be this low we need a large temporal extend of the lattice, this leads to a negligible effective gauge coupling i.e. $\lambda(\beta=2.5,N_t=200)\sim 1\cdot 10^{-15}$. We therefore end up with a completely fermionic partition function. The only leftovers from the Yang-Mills part of the original theory come from gauge corrections to the effective fermion couplings, which are the same as in $SU(3)$ and amount to replacing $h$ in (\ref{ferm_coupling}) by
\begin{equation}
	h = \exp \left[ N_t \left(a \mu + \ln 2 \kappa + 6 \kappa^2 \frac{u-u^{N_\tau}}{1-u} + \, \cdots\right) \right] \; . \label{h1}
\end{equation}
Now we can piece everything together to get the effective action for the cold and dense regime and one flavour:
\begin{align}
&-S_\text{eff}= \sum_{\vec x} \log (1+hL_{\vec x} +h^2)^2 \notag \\
&-2 h_2 \sum_{\vec x, i} \tr \frac{h W_{\vec x}}{1 +h W_{\vec x}} \tr \frac{h W_{\vec x+i}}{1+h W_{\vec x+i}} \notag \\
&+2 \frac{\kappa^4 N_t^2}{N_c^2} \sum_{\vec x, i} \tr \frac{h W_{\vec x}}{(1 +h W_{\vec x})^2} \tr \frac{h W_{\vec x+i}}{(1+h W_{\vec x+i})^2} \notag \\
&+\frac{\kappa^4 N_t^2}{N_c^2} \sum_{\vec x, i, j} \tr \frac{h W_{\vec x}}{(1 +h W_{\vec x})^2} \tr \frac{h W_{\vec x-i}}{1+h W_{\vec x-i}} \tr \frac{h W_{\vec x-j}}{1+h W_{\vec x-j}} \notag \\
&+2\frac{\kappa^4 N_t^2}{N_c^2} \sum_{\vec x, i, j} \tr \frac{h W_{\vec x}}{(1 +h W_{\vec x})^2} \tr \frac{h W_{\vec x-i}}{1+h W_{\vec x-i}} \tr \frac{h W_{\vec x+j}}{1+h W_{\vec x+j}} \notag \\
&+\frac{\kappa^4 N_t^2}{N_c^2} \sum_{\vec x, i, j} \tr \frac{h W_{\vec x}}{(1 +h W_{\vec x})^2} \tr \frac{h W_{\vec x+i}}{1+h W_{\vec x+i}} \tr \frac{h W_{\vec x+j}}{1+h W_{\vec x+j}} \notag \\
  &+\kappa^4 N_t^2 \sum_{\vec x,i} \frac{h^4}{(1+h L_{\vec x}+h^2)(1+h L_{\vec x+i}+h^2)} \; , \label{eff_action_kappa4}
  \end{align}
where the second fermion coupling $h_2$ is defined as
\begin{equation}
	h_2= \frac{\kappa^2 N_t}{N_c}\left[1+2 \frac{u-u^{N_t}}{1-u}+ \, \cdots \right] \; . \label{h2}
\end{equation}
Note that the combined strong-coupling hopping expansion is defined as an expansion in $\kappa^m u^n$, and given here to the order $m+n=4$. Unlike the one-point fermion coupling $h$ in (\ref{h1}), a partial resummation of $h_2$ is not possible at this order \cite{Langelage}. We furthermore dropped all terms proportional to $\bar h = h(-\mu)$ in Eq.~(\ref{eff_action_kappa4}) as well as terms that are subleading in $N_t$. This implies that the form of the effective action in (\ref{eff_action_kappa4}) is valid only for sufficiently large $\mu$ and $N_t$.

\subsubsection{Effective Action for $N_f$ Flavors} 
\label{ssub:effective_action_for_two_flavors}
In the theory with $N_f$ quark flavors we have to introduce fermion determiants for each of them in Eq.~\eqref{eff_action}. When they are degenerate this simply amounts to taking the power $N_f$ of the fermion determinant. As in the one-flavor case we split the fermion determinant into a static and a kinetic part and do the hopping expansion analogous to that in Eqs.~\eqref{determinant} and \eqref{det_kin}:
\begin{align}
\det [D]^{N_f} =& (\det [D_\text{stat}] \det[D_\text{kin}])^{N_f}   \\
& \hskip - 1cm = \det [D_\text{stat}]^{N_f} \exp \Big[ - N_f \,  \tr PM - N_f \,  \tr PPMM  \notag \\
  &- \frac{N_f}{2} \,  \tr PMPM +\mathcal{O}(\kappa^6) \Big]  \notag \\
& \hskip -1cm = \det [D_\text{stat}]^{N_f}   \Big[1 -N_f\,  \tr PM -N_f \,  \tr PPMM  \notag \\
  &- \frac{N_f}{2} \, \tr PMPM + 
  \frac{N_f^2}{2} \,  (\tr PM)^2 +\mathcal{O}(\kappa^6) \Big]  \; .  \notag 
\end{align}
This leads to the effective action for $N_f$ degenerate flavors:
\begin{align}
  &\hskip -.16cm -S_\text{eff}=N_f \sum_{\vec x} \log (1+hL_{\vec x} +h^2)^2 \\
&\hskip -.16cm -2N_f h_2 \sum_{\vec x, i} \tr \frac{h W_{\vec x}}{1 +h W_{\vec x}} \tr \frac{h W_{\vec x+i}}{1+h W_{\vec x+i}} \notag \\
&\hskip -.16cm +2N_f^2 \frac{\kappa^4 N_t^2}{N_c^2} \sum_{\vec x, i} \tr \frac{h W_{\vec x}}{(1 +h W_{\vec x})^2} \tr \frac{h W_{\vec x+i}}{(1+h W_{\vec x+i})^2} \notag \\
&\hskip -.16cm +N_f \frac{\kappa^4 N_t^2}{N_c^2} \sum_{\vec x, i, j} \tr \frac{h W_{\vec x}}{(1 +h W_{\vec x})^2} \tr \frac{h W_{\vec x-i}}{1+h W_{\vec x-i}} \tr \frac{h W_{\vec x-j}}{1+h W_{\vec x-j}} \notag \\
&\hskip -.16cm +2N_f \frac{\kappa^4 N_t^2}{N_c^2} \sum_{\vec x, i, j} \tr \frac{h W_{\vec x}}{(1 +h W_{\vec x})^2} \tr \frac{h W_{\vec x-i}}{1+h W_{\vec x-i}} \tr \frac{h W_{\vec x+j}}{1+h W_{\vec x+j}} \notag \\
&\hskip -.16cm +N_f\frac{\kappa^4 N_t^2}{N_c^2} \sum_{\vec x, i, j} \tr \frac{h W_{\vec x}}{(1 +h W_{\vec x})^2} \tr \frac{h W_{\vec x+i}}{1+h W_{\vec x+i}} \tr \frac{h W_{\vec x+j}}{1+h W_{\vec x+j}} \notag \\
&\hskip -.16cm +(2N_f^2-N_f)\kappa^4 N_t^2 \sum_{\vec x,i} \frac{h^4}{(1+h L_{\vec x}+h^2)(1+h L_{\vec x+i}+h^2)} \; \notag . \label{eff_action_kappa4_Nf2}
\end{align}

\subsection{Symmetries and Spectrum of two-color QCD}
One of the unique features of two-color QCD is the fact that the fundamental representation of $SU(2)$ is pseudo-real, i.e.~for vanishing chemical potential, quarks and anti-quarks belong to equivalent representations. This results in an enlarged flavor symmetry. In the continuum theory with $N_f$ degenerate quark flavors the symmetry group is $SU(2N_f)$ which contains the usual chiral $SU(N_f)_L \times SU(N_f)_R \times U(1)_B$ symmetry as a subgroup, see e.g.~\cite{VonSmekal2012}. For $N_f=2$ continuuum flavors the enlarged flavor symmetry is $SU(4)$ and the chiral symmetry breaking pattern is $SU(4)\to Sp(2)$ which is locally isomorphic to $SO(6)\to SO(5)$. Dynamical chiral symmetry breaking then leads to five Goldstone bosons in the chiral limit: the three pions and a scalar diquark/anti-diquark pair. For a small finite quark mass $m_q$ these diquarks are thus pseudo-Goldstone bosons degenerate with the pions. As such they also are the lightest baryonic excitations in two-color QCD with bosonic baryons. 
The enlarged flavor symmetry is also broken by a finite baryon-chemical potential $\mu_B = 2\mu>0$, as $SU(4)\to SU(2)_L \times SU(2)_R \times U(1)_B$. With both, $m_q >0$ and $\mu_B>0$, one is left with the usual isospin and baryon number symmetries, $SU(2)_V \times U(1)_B$. The two-color analogue of the nuclear-matter transition is Bose-Einstein condensation of the scalar diquarks, where a non-vanishing diquark condensate $\langle qq\rangle $ dynamically breaks the baryon number $U(1)_B$. Because this transition is of second order and hence continuous, for $T=0$ it must occur at $\mu_B = m_d$, i.e. $\mu_c = m_d/2 = m_\pi/2$. In particular, there is no shift in $\mu_c$ by binding energy as in the liquid-gas transition of nuclear matter in QCD (where $3\mu_c = m_B - \epsilon = 923 $ MeV for a nucleon mass $m_B= 939$~MeV and a binding energy per nucleon of $\epsilon = 16$ MeV). 

From these general symmetry considerations,  one would naively predict for $N_f=1$ a chiral symmetry breaking pattern $SU(2)\to U(1)_B$ with a scalar diquark/anti-diquark pair as the two pseudo-Goldstone bosons. However, it turns out that one cannot construct a totally antisymmetric color-singlet pair of quarks in the scalar channel with only one flavor. There is therefore no diquark condesate for $N_f=1$ either and baryon number remains unbroken, at least in the confined phase.

Since the present approach is valid only for very heavy quarks, however, there is no approximate chiral symmetry in the first place.\footnote{At least it means that we don't have to worry about the poor chiral properties of Wilson fermions for which everything except the $SU(2)_V\times U(1)_B$ is broken explicitly by the Wilson term.} Nevertheless, for $N_f=2$ there is a scalar diquark which still remains exactly degenerate with the pion in the vacuum. It is therefore straightforward to calculate its mass in the combined strong coupling and hopping expansion. The result is then the same as for the pion, 
\begin{equation}
 a m_d= - 2 \ln(2\kappa) - 6 \kappa^2 -24 \kappa^2 \frac{u}{1-u} + 6 \kappa^4 +\, \cdots \; , \label{mass}
\end{equation}
with corrections of the order $n+m =6$.
For comparison, we can estimate the quark mass by assuming $m_q = \mu - T\ln h$ as at leading order, but now with the strong-coupling and hopping expansion corrections included in $h$ which are formally the same as in $SU(3)$ and available to the order $m+n = 7$ \cite{Fromm2012}. This corresponds to extracting the quark mass from the static propagator $D_\text{stat}^{-1}$ at this order and yields,
\begin{align}
  a m_q =& - \ln(2\kappa) - 6 \kappa^2 \frac{u}{1-u} + 48 \kappa^4 u  (1-u)  \notag \\
  &- 24 \kappa^2 u \, (u^4  + \kappa^2 u^2 -  \kappa^4 ) + \, \cdots \; . \label{quarkmass}
\end{align}
It amounts to a small binding energy for the scalar diquark beyond leading order, in lattice units,
\begin{equation}
  2 a m_q - am_d = 6\kappa^2 \frac{1+u}{1-u} - 6\kappa^4 (1-8 u(1-u)) + \, \cdots \; . \label{dbind}
\end{equation}
We reiterate that this scalar diquark does not exist in the one-flavor case.

\subsection{Leading-Order Mean-Field Density}
\label{Background_LOMF}
We can get some insight into the model by looking at the static part of the fermion determinant in the strong coupling limit $\lambda=\beta=0$. Then, the partition function factorizes,
\begin{align}
	Z(\beta=0)= \left( \int dW \; [1+h L_{\vec x}+h^2]^{2N_f} \right. \\
	 \times [1+\bar h L_{\vec x}+\bar{h}^2]^{2N_f} \text{\huge)}^{N_s^3} \notag \; .
\end{align}
For  $T \rightarrow 0$ at a finite chemical Potential $\mu$ we have $\bar h \rightarrow 0$. Within a mean-field description of the Polyakov loop this partition function then simplifies to
\begin{align}
 Z= [1+2 h \tilde L+h^2]^{2 N_f N_s^3} \; ,
\end{align}
where a factor of two was inserted because the mean-field Polyakov loop $\tilde L$ here is normalized to assume values within $[-1,1]$. The quark number density then follows as
\begin{align}
n &= \frac{T}{V} \frac{\partial}{\partial \mu} \ln Z \; , \notag \\
a^3 n&= 4 N_f \frac{1+\tilde Le^\frac{m_q-	\mu}{T}}{1+2\tilde Le^\frac{m_q-	\mu}{T}+e^\frac{2(m_q-	\mu)}{T}} \; . \label{mean-field_dens}
\end{align}
The same expression can be derived from a Polyakov-Quark-Meson-Diquark model  \cite{Strodthoff:2013cua} for very heavy quarks. For  $\tilde L=1$ it simply reduces to the zero-momentum occupation number of the Fermi-Dirac distribution. In the deconfined phase it thus describes a free gas of heavy quarks. We will see below that for small but finite Polyakov-loop expectation values $\tilde L > 0$, and small $T$,  Eq.~(\ref{mean-field_dens}) due to imperfect confinement behaves as a quark gas, suppressed by $\tilde L$, up to some critical chemical potential $\mu_c(T)$ where it starts to reflect the behavior of a diquark gas with $m_d = 2 m_q$ at this leading order. Explicitly, consider $\mu < m_q$ and hence $ x \equiv \exp\{(m_q -\mu)/T\} > 1$. We can then distinguish two regimes:
\begin{equation}
  \frac{a^3n }{4 N_f} \to \left\{ \begin{array}{ll}  e^{(\mu_B - m_d)/T}\; , & \tilde L \, x \ll 1 \; ,  \\[4pt]
 \tilde L\,  e^{(\mu - m_q)/T}\; , & \tilde L\, x \gg 1 \; .
  \end{array} \right. \label{MF_limits}
\end{equation}
The transition occurs at around $\mu_c \sim m_q + T \ln \tilde L \, $.

 \begin{figure}[t]
 \includegraphics[width=.9\linewidth]{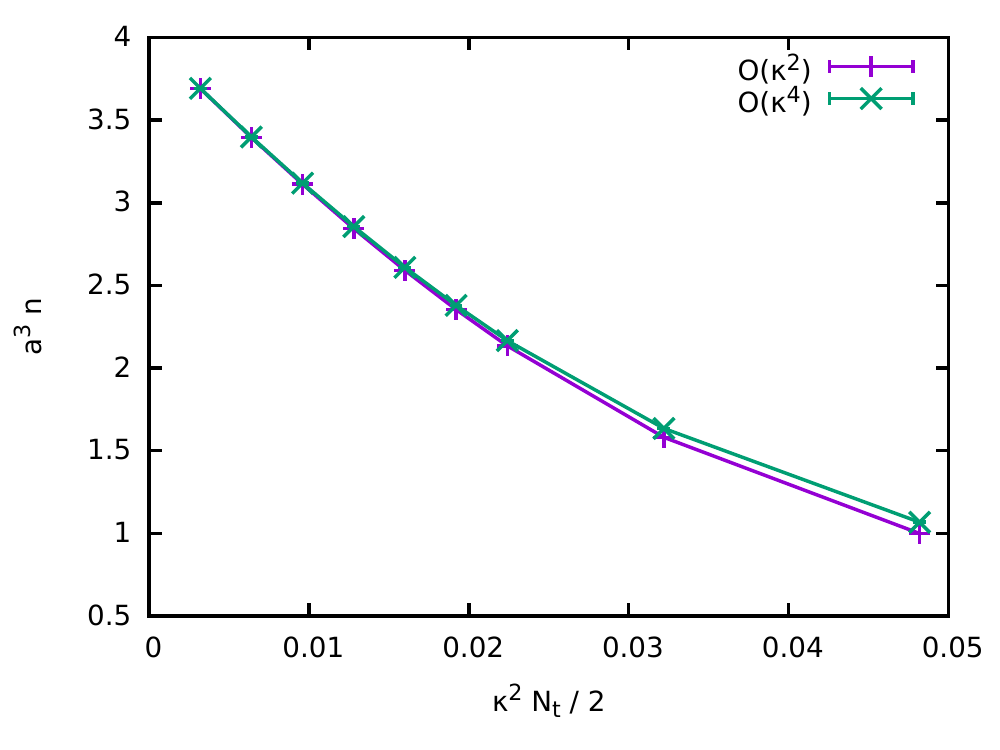}%
 \caption{Comparison of quark densities evaluated at the orders $\kappa^2$ and $\kappa^4$, both for $\kappa = 0.00802 $, $\beta=2.5$, $\mu= m_d/2$ and $N_t$ ranging between 100 and 2400, including gauge corrections. \label{convergence}}
 \end{figure}
\section{Results}
In order to assess the range of validity of the effective theory we need to test the convergence properties of the hopping series for various parameter choices. As can be seen from Equation (\ref{eff_action_kappa4}) the relevant expansion parameter for the effective theory is $\kappa^2 N_t/N_c$. Because we are interested in very low temperatures $T=1/a N_t$, we need lattices of large extend $N_t$ in the temporal direction, especially as we go to smaller lattice spacings $a$. Therefore our hopping parameter $\kappa$ needs to be sufficiently small. This implies that one can reach smaller quark masses at higher temperatures and vice versa. To check the convergence of the hopping expansion we compare expectation values for the densities $a^3n$ from simulations including corrections up to $\mathcal{O}(\kappa^2)$ and $\mathcal{O}(\kappa^4)$ for different values of the expansion parameter $\frac{\kappa^2 N_t}{N_c}$. Figure \ref{convergence} shows a plot of the two densities for $\mu = m_d/2$ at $\beta=2.5$, including gauge corrections. The two agree reasonably well for all values of $\kappa^2 N_t/{2}$ up to slightly above 0.02.

\subsubsection*{Scale Setting and Units} 
Assuming that quarks as heavy as those used here only have a negligible influence on the running of the coupling, we use the non-perturbative $\beta$-function from \cite{Smith2013}
\begin{equation}
	\frac{1}{a \sqrt{\sigma}}= \exp \left( \frac{\beta-d}{b}\right) \; , \label{scale}
\end{equation}
with the parameters
\begin{equation}
	d=1.98(1) \; , \hspace{0.5cm} b=0.305(6) \; .
\end{equation}
In lack of phenomenological input we somewhat arbitrarily use the typical $\sqrt{\sigma}=440$ MeV also for the string tension of two-color QCD.
With Eqs.~\eqref{scale} and  \eqref{mass} we are then able to assign physical scales to our systems. In all our simulations the diquark mass from Eq.~(\ref{mass}) is adjusted in this way to $m_d=20$ GeV, and the temperatures range between $T= 3.454$ MeV and $9$ MeV. On our finest lattice with $\beta = 2.5$, corresponding to $a=0.0810 $~fm, this amounts to $\kappa = 0.00802123$ and $N_t$ values between $ 269$ and $700$ such that $\kappa^2 N_t/2 $ lies within $0.0087$ and $0.0225$, and hence within the range of validity of the hopping expansion, cf.~Fig.~\ref{convergence}. The parameters for the coarser lattices with $\beta$ values down to 2.4 lead to even smaller values for the expansion parameter.

\begin{figure}
 \includegraphics[width=.9\linewidth]{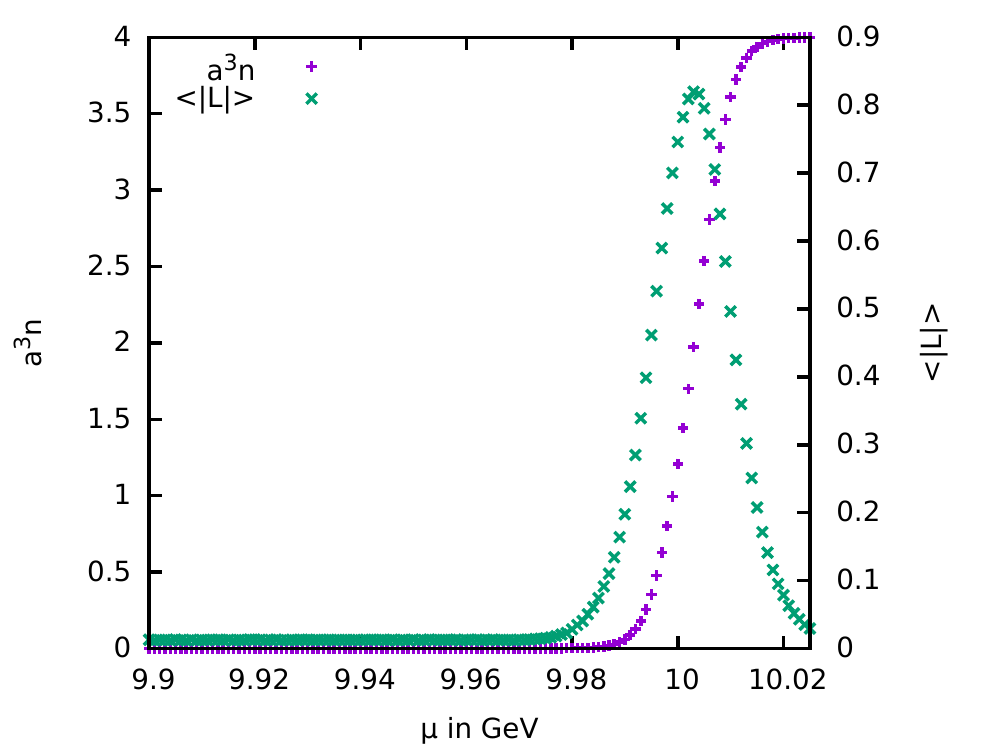}%
 \caption{Quark density  $a^3n$ in lattice units and Polyakov-loop expecation value $\langle |L| \rangle$, both over $\mu$, with simulation parameters $\beta=2.5$, $\kappa=0.00802$, leading to $m_d=20$~GeV, $N_s=16$, $N_t=484$, corresponding to $T=5$ MeV, and $N_f=1$. \label{linear_dens_poly}}
 \end{figure}

\subsection{Results for \boldmath $N_f=1$}
In this subsection we first present our numerical results for the Polyakov loop and the quark density in the effective theory at low temperatures with one quark flavor, $N_f=1$.
	
As an indication of deconfinement at high density and low temperature we plot the the Polyakov-loop expectation value in Fig.~\ref{linear_dens_poly}, where $\langle |L| \rangle $ stands for the usual expectation value of the modulus of the volume averaged 
\begin{equation} 
L \equiv \frac{1}{V} \sum_{\vec x} L_{\vec x}\; .
\end{equation}
Because of the presence of dynamical quarks, even for vanishing net-baryon density, the Polyakov loop will have a small but nonzero expectation value $\langle |L| \rangle > 0$. For the parameters of Fig.~\ref{linear_dens_poly} its $\mu=0$ value is $\langle |L| \rangle = 0.012$. For aspect ratios as the one considered here, with $N_t/N_s \approx 30$ in this example, this value is determined by the finite spatial volume. It is therefore basically temperature independent. It furthermore also remains constant in $\mu$ until just below the onset of the density near $m_d/2$ because the temperature of $ T= 5$~MeV here is so low that no baryonic degrees of freedom are being excited as long as the baryon chemical potential $2\mu$ stays well below the gap in the baryon spectrum. From Fig.~\ref{linear_dens_poly} it appears however that $\langle |L| \rangle$ starts to rise from its $\mu=0$ value before the onset of the density so that we can not distinguish baryon density from quark density here. As we will discuss in the next subsection, however, there is a very small diquark contribution in the density above $\mu_c \approx 9.96$ GeV at this temperature which is not visible on the linear scale of Fig.~\ref{linear_dens_poly} but where the Polyakov loop still has its zero-density expectation value. We will provide some evidence that this diquark-density onset might stay below the deconfinement transition when we extrapolate both to $T=0$ for $N_f=2$ with scalar diquark below. 

\begin{figure}
 \includegraphics[width=.9\linewidth]{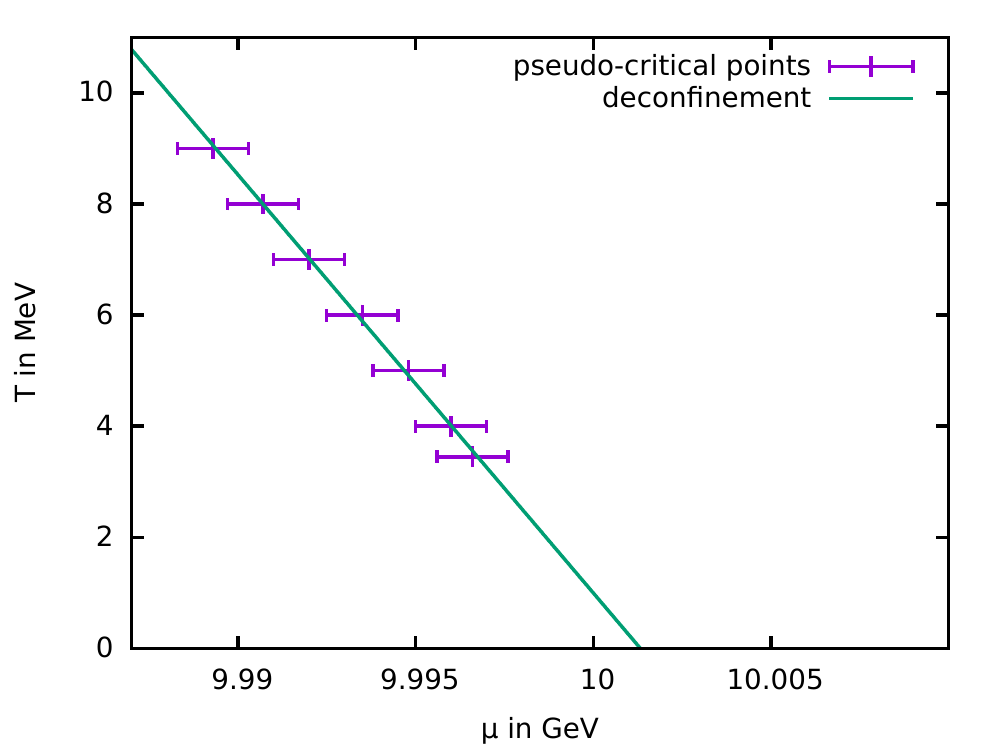}%
 \caption{The pseudo-critical line for the deconfinement transition from simulations with the same parameters as in Fig.~\ref{linear_dens_poly} except for different $N_t$, corresponding to $T = 9,\, 8, \dots 3.454$~MeV, and
   a linear extrapolation to $T=0$.
   \label{pc_poly}}
 \end{figure}

At larger values of the chemical potential the quark density saturates at $a^3 n = 2 N_c N_f$, the maximum number of quarks per site due to the Pauli principle, as in the effective theory for heavy quarks in $SU(3)$ \cite{Fromm:2012eb}. This behavior which is a lattice artifact has previously also been observed in finite density simulations of two-color \cite{Hands2006} and $G_2$-QCD \cite{Maas2012}. This saturation leads to an effective quenching of the quarks and hence the  Polyakov loop decreases again as it is approached. We also analyzed the Polyakov-loop susceptibility and found no increase of its rather broad maximum with the lattice volume hence indicating a smooth cross-over behavior rather than a deconfinement phase transition in the infinite volume limit. The pseudo-critical chemical potential $\mu_{pc}$ from the inflection point of $\langle |L| \rangle$ along the $\mu$ axis is shown in Fig.~\ref{pc_poly}. It coincides with the point where $\langle |L| \rangle$ reaches half its maximum value. By determining this $\mu_{pc}$ for different temperatures we obtain a pseudo-critical line for the deconfiment transition at low temperature which can be  extrapolated to $T=0$ as shown in Fig.~\ref{pc_poly}.  We can see that the pseudo-critical line terminates at $\mu_{pc}$ slightly above $10$~GeV which corresponds to half the scalar-diquark mass from  Eq.~(\ref{mass}) with the parameters used here.

	 \begin{figure}
 \includegraphics[width=.9\linewidth]{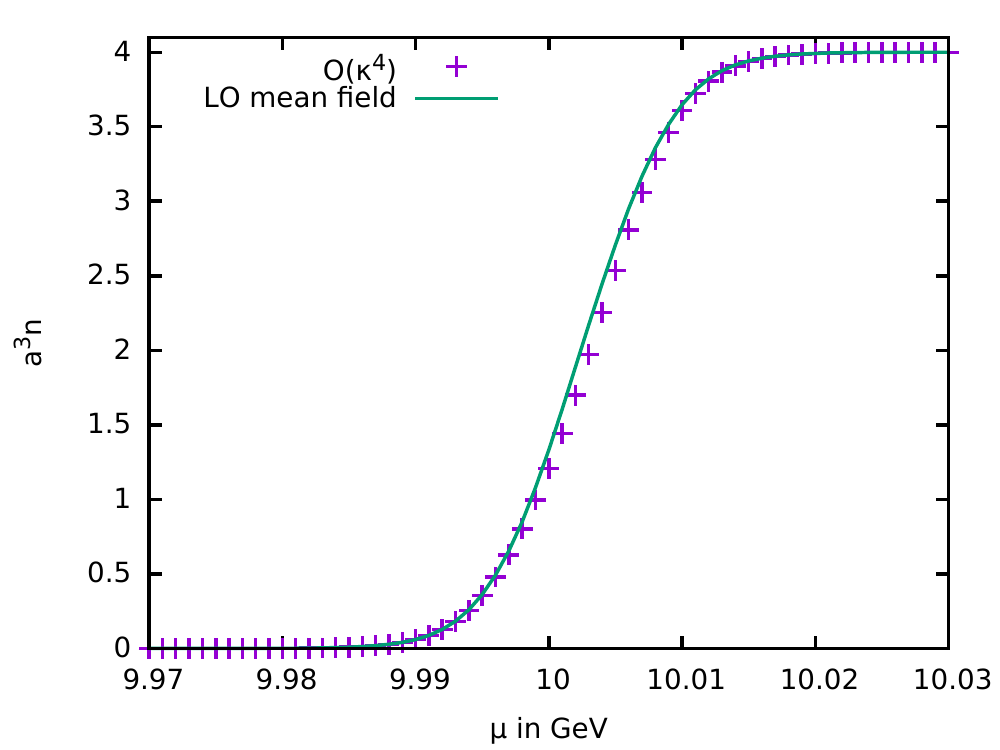}%
 \caption{Comparison of the quark density at order $\kappa^4$  from Fig.~\ref{linear_dens_poly} to the leading-order formula in Eq.~(\ref{mean-field_dens}) with $\tilde L$ replaced by the corresponding data for $\langle |L|\rangle/2 $ from Fig.~\ref{linear_dens_poly}, $T=5 $~MeV. \label{vgl_LOMF}}
\end{figure}
 
We conclude this subsection by discussing in some more detail the quark-number density $n$ defined by
\begin{equation}
	n= \frac{T}{V} \frac{\partial}{\partial \mu} \ln Z \; .
\end{equation}
As mentioned above, Fig. \ref{linear_dens_poly} shows the strong increase in the density in lattice units $a^3n$ at a value of the quark chemical potential around $\mu=m_d/2$, from where on it rapidly grows to its saturation value 
with each lattice site fully occupied by $2 N_c N_f=4$ quarks.
This transition is described qualitatively well by the leading-order mean-field formula in Eq.~(\ref{mean-field_dens}). This can be seen in Fig.~\ref{vgl_LOMF} where we compare the data for the quark density at order $\kappa^4$ from Fig.~\ref{linear_dens_poly} to the leading order form in Eq.~(\ref{mean-field_dens}) with the corresponding $am_q = - \ln(2\kappa) $ for the strong-coupling limit, and with $\tilde L $ replaced by $\langle |L| \rangle/2$, i.e., using the $\mu$-dependent data for the Polyakov-loop expectation value of Fig.~\ref{linear_dens_poly} in the mean-field approximation. 

\begin{figure} 
 \includegraphics[width=.9\linewidth]{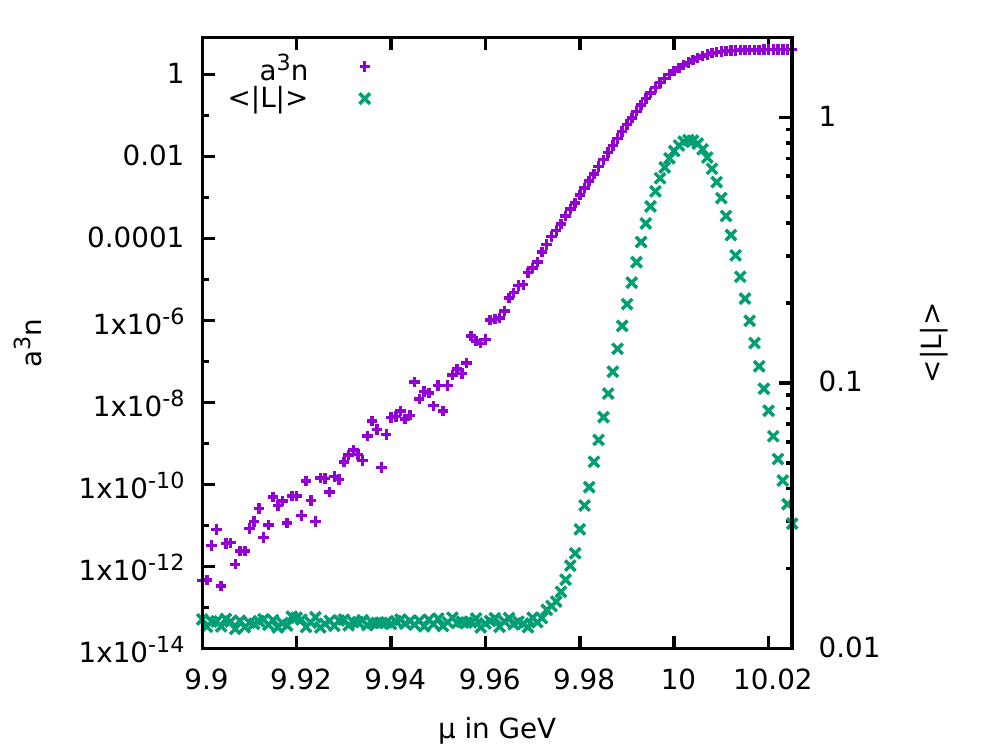}%
 \caption{Logarithmic plot of the density in lattice units $a^3n$ and the Polyakov loop $\langle |L| \rangle$ as a function of $\mu$ in physical units  at $T=5$ MeV with the same parameters as in Fig.~\ref{linear_dens_poly}. \label{log_dens_poly}}
 \end{figure}

To resolve the differences we need to have a closer look at the behavior of the chemical-potential and temperature dependence of the quark density, especially in the region where $\tilde L \, x \sim 1$ with $x= \exp\{(m_q-\mu)/T\}$ as defined in Sec.~\ref{Background_LOMF} above. Fig.~\ref{log_dens_poly} shows a logarithmic plot of the density and the Polyakov loop of Fig.~\ref{linear_dens_poly}.

We observe two different regimes of exponential increase before the density approaches its saturation value. They are separated by a kink in the logarithmic plot, here at $\mu_c \approx 9.96$~GeV, where the Polyakov-loop still is constant at its $\mu=0$ expectation value,  $\langle |L|\rangle/2 \approx 0.006$.

\begin{figure}
 \includegraphics[width=.9\linewidth]{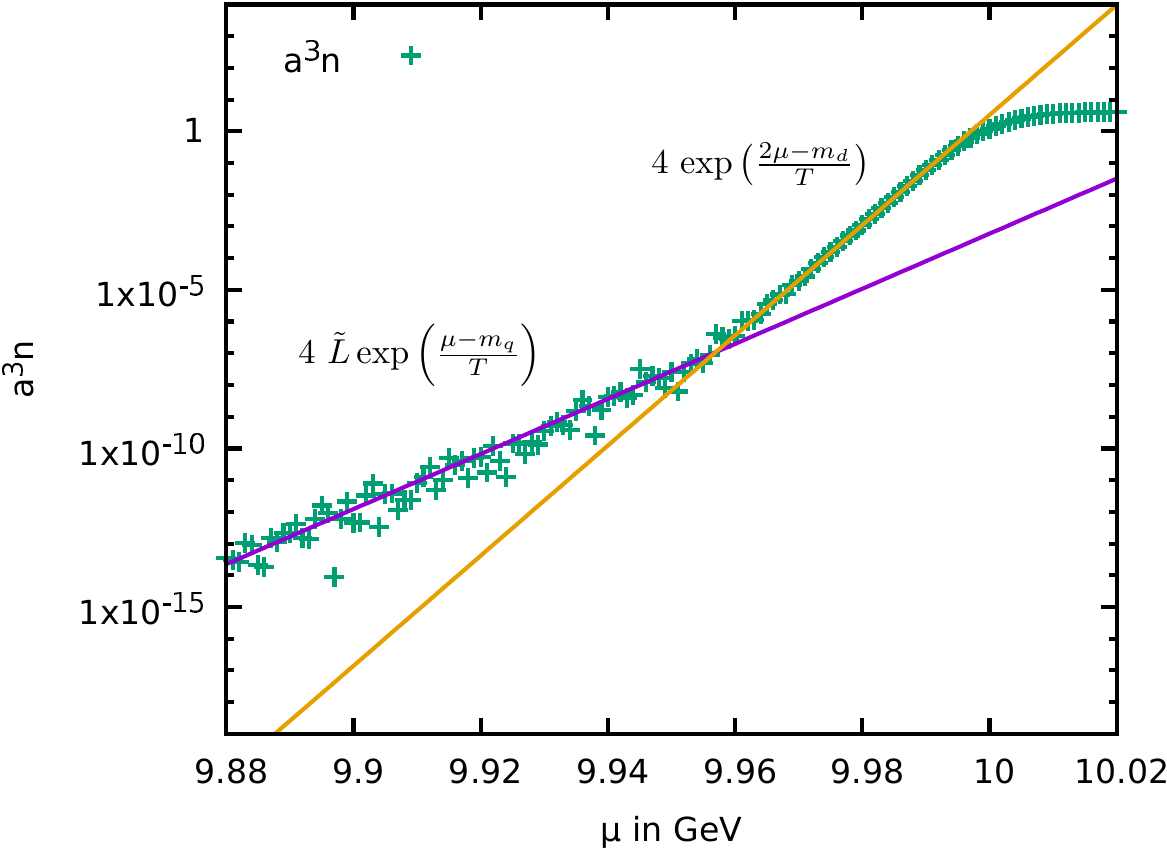}%
 \caption{Logarithmic plot of the density in lattice units $a^3n$ at $T=5$ MeV compared to one-parameter fits using $\tilde L$ for $\mu < 9.96$~GeV and $m_d$ for $\mu > 9.96$~GeV, see text. \label{slopes}}
 \end{figure}

The two regimes correspond to the two limits in (\ref{MF_limits}) of the leading-order mean-field density, Eq.~(\ref{mean-field_dens}). This is demonstrated in Fig.~\ref{slopes} where we compare the density to two corresponding fits:

When we fit the data in the region of the second exponential increase, for $\mu $ values between 9.96~GeV and 9.99~GeV, to
\begin{equation}
  a^3 n = 4 N_f \exp\{(2\mu - m_\mathrm{fit})/T \} \label{dfit}
\end{equation}
with a single parameter $m_\mathrm{fit}$ we obtain for $N_f=1$,
\begin{equation}
  m_\mathrm{fit} = 20.0045(5)\, \mbox{GeV}\, . \label{mfitNF1}
\end{equation}
For comparison, with the same lattice parameters the quark mass from Eq.~(\ref{quarkmass}) becomes $m_q^{(0)} = 10.0024$~GeV at the leading order $n+m=0$, 
$m_q^{(4)} = 10.0014 $~GeV at the order $n+m = 4$, and $ m_q^{(7)} = 10.0013 $~GeV at the order $n+m = 7$. Therefore, the fit parameter $m_\mathrm{fit} $ is consistent with $2 m_q^{(0)}$ but slightly larger than $2m_q = 20.0028(2)$ at the same order $n+m=4$ (with an error of the size of the higher-order corrections up to $ n+m =7$ as given explicitly in Eq.~(\ref{quarkmass})). It is larger than the corresponding scalar diquark mass, $m_d= 20$~GeV from Eq.~(\ref{mass}), which might simply reflect the fact this scalar diquark does not exist for $N_f = 1$. A daring interpretation would be that we see a heavier diquark mass here, such as that of an axial-vector diquark which can exist also for $N_f=1$.

To test this, we have done the same analysis with the same parameters also for two flavors (see the next subsection). The same fit to the form in (\ref{dfit}) then yields for $N_f=2$,
\begin{equation}
  m_\mathrm{fit} = 19.9986(10)\, \mbox{GeV}\, ,  \label{mfitNF2}
\end{equation}
which is now indeed very close to the scalar diquark mass $m_d$ and significantly smaller than $2m_q$.
In order to quantitatively describe this regime of exponential increase we therefore have to replace $2m_q$ by $m_d$ in the leading-order mean-field formula for the density in  Eq.~(\ref{mean-field_dens}). This indicates that matter on this side of the kink consists of diquark excitations.

The first exponential increase, for the $\mu$ values below $9.96$~GeV, is described by
\begin{equation}
  a^3n = 4N_f\,  \tilde L \, \exp\{(\mu-m_q)/T\} \; . \label{qfit}
\end{equation}
In this case we use $m_q = 10.0014$~GeV for the quark mass at this order and fit the data via $\tilde L$ as the free parameter. For the one-flavor data of Fig.~\ref{slopes} this leads to $\tilde L $ of the order of $10^{-4} $, however, with a very large uncertainty. It determines the precise value of the onset $\mu_c$ of the diquark density, by the intersection point of the two different exponential fits (\ref{dfit}) and (\ref{qfit}), as    
\begin{equation}
  \tilde L = \exp\{(\mu_c - m_d +m_q)/T\} \; .
\end{equation}
With $m_d = m_\mathrm{fit}$ from (\ref{mfitNF1}) for the $N_f=1$ data in Fig.~\ref{slopes}, for example, this leads to values between $\tilde L = 8\cdot 10^{-5}$ for $\mu_c = 9.956$~GeV and $\tilde L = 1.2\cdot 10^{-4} $ for  $\mu_c = 9.958$~GeV. In any case, it is much smaller than the zero-density value of $\langle |L|\rangle/2 \approx 0.006$. Instead we observe that it is more consistent with the expectation value of the local Polyakov loop $L_{\vec x}$. Its expectation value is extracted from the per-site probability distribution $P(L_{\vec x})$ which we otain by histograming the local Polyakov-loop variable $L_{\vec x}$ as in \cite{Smith2013}. At $T=5$ MeV, with $\beta=2.5$ and $\kappa = 0.00802$, we obtain for this observable a zero-density value of about $\langle L_{\vec x} \rangle\sim 10^{-4}$ instead of  $\langle |L| \rangle \approx 0.012$ for the modulus of the volume-averaged Polyakov loop. This suggests that one should use the local Polyakov-loop expectation value $\langle L_{\vec x}\rangle$ in mean-field approximations as Eq~(\ref{mean-field_dens}). Like $\langle |L| \rangle$ it is independent of the chemical potential below the deconfinement crossover at $\mu_{pc}$. And as soon as the Polyakov-loop starts to rise from its constant zero-density expectation value the two agree well within the errors. It is only the residual small value at vanishing net-baryon density due to imperfect confinement in a finite volume in which the two differ simply because it takes longer for the modulus to vanish than the local Polyakov loop in the infinite-volume limit for  $\mu< \mu_{pc} $. This difference is only relevant at densities in lattice units below $10^{-4}$ and hence not visible on the linear scale of Fig.~\ref{vgl_LOMF} above.

\begin{figure}
 \includegraphics[width=.9\linewidth]{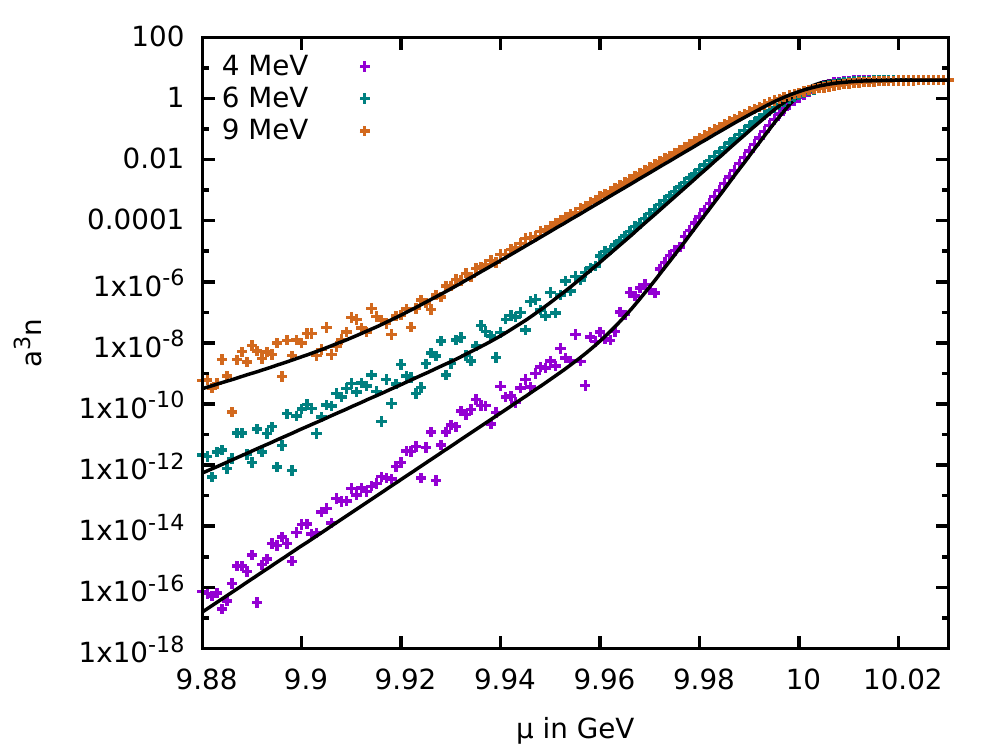}%
 \caption{Comparison of the measured density with the leading-order mean-field density with $\tilde L = \langle L_{\vec x}\rangle /2$ at $m_d=20$ GeV, $\beta=2.5$, $\kappa = 0.00802$, and different temperatures. \label{nofit}}
\end{figure}

In fact, using the local Polyakov-loop expectation value $\langle L_{\vec x}\rangle $ for various temperatures in the leading-order mean-field formula for the quark density, Eq.~(\ref{mean-field_dens}), describes the data especially also in the low-density region around the diquark-density onset at $\mu_c$ very well as can be seen in Fig.~\ref{nofit}. These are not fits. We simply use Eq.~(\ref{mean-field_dens}) with $\tilde L = \langle L_{\vec x}\rangle/2 $ here to describe the quark density over the whole range of temperatures  we have investigated. It describes the imperfect statistical confinement of quarks for $\mu$ below $\mu_c = m_d-m_q + T \ln(\langle L_{\vec x}\rangle/2)$, an ensemble of diquarks above $\mu_c$, and quark matter with lattice saturation for $\mu$ larger than $\mu_{pc}$ where $2\tilde L = \langle L_{\vec x} \rangle = \langle |L|\rangle$ as in Fig.~\ref{vgl_LOMF}, all at the same time.

\subsection{Results for $N_f=2$}
In this subsection we discuss the results from the effective theory with $N_{f}=2$ degenerate quark flavors in somewhat more detail. In particular, we describe how well the onset of the diquark density at $\mu=\mu_c$  agrees with the scalar diquark mass $m_d$ from Eq.~(\ref{mass}) which we know that it exists for $N_f=2$.

\begin{figure}
 \includegraphics[width=.9\linewidth]{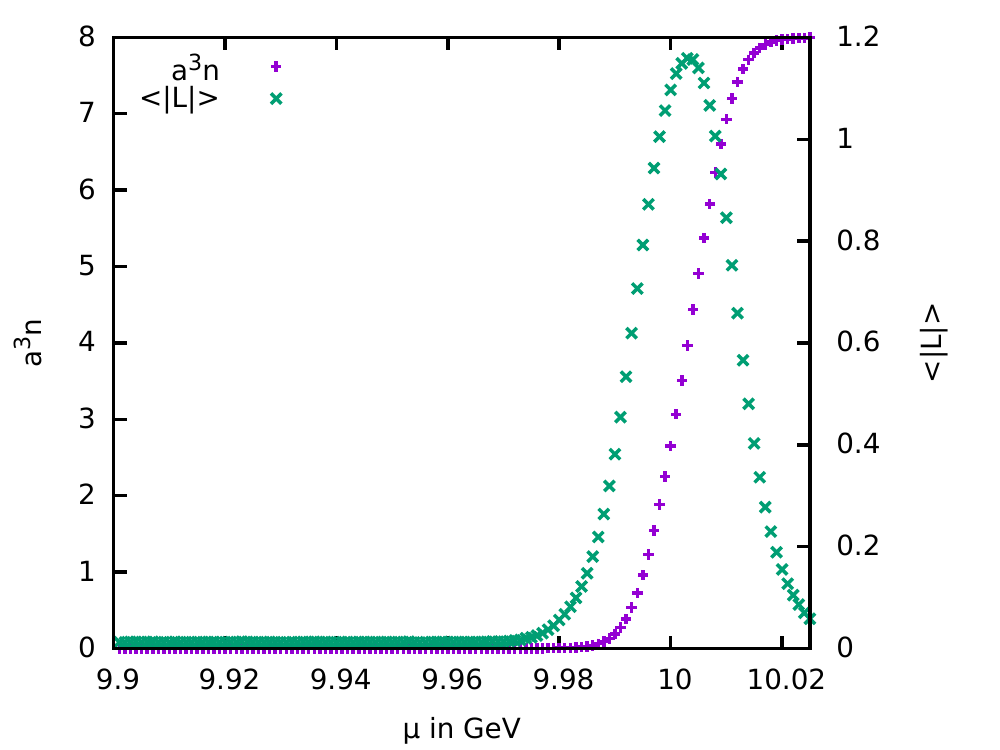}%
 \caption{Quark density in lattice units and Polyakov-loop expectation value $\langle |L| \rangle$, both over  $\mu$, with the same simulation parameters as in Fig.~\ref{linear_dens_poly}, but for $N_f=2$ here. \label{linear_dens_poly_Nf2}}
 \end{figure}

As a first example in Fig.~\ref{linear_dens_poly_Nf2} we present the $N_f=2$ results for the quark density and the Polyakov loop on our finest lattice, with $\beta = 2.5$, $\kappa = 0.00802123$, $N_s = 16$ and $N_t=484$ corresponding to $m_d=20$~GeV and $T= 5$~MeV at a lattice spacing of $a= 0.081$~fm as in the previous subsection for the one-flavor case. Both observables show the same qualitative behavior as for $N_f=1$ before.

\begin{figure}
 \includegraphics[width=.9\linewidth]{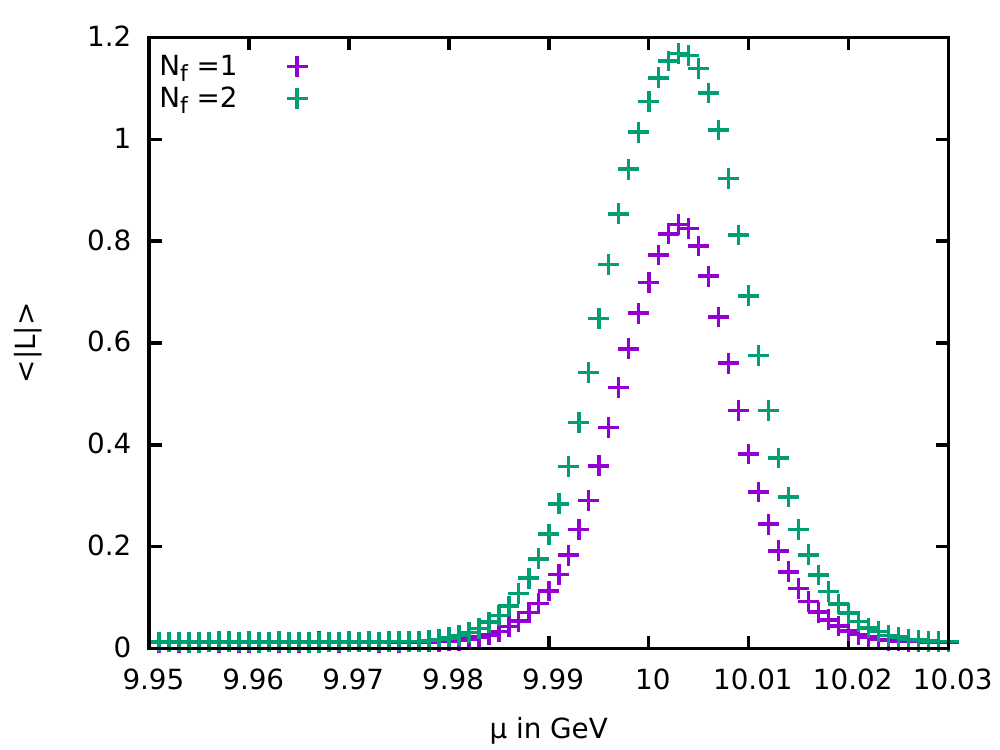}%
 \caption{Comparison of the Polyakov loop $\langle |L| \rangle$ for $N_f=1$ and $2$  at $T=5$ MeV with parameters as in Figs.~\ref{linear_dens_poly} and \ref{linear_dens_poly_Nf2}. \label{vlg_poly}}
 \end{figure}

The most obvious differences between $N_f=1$ and $2$ are the different saturation densities given by $a^3n_\text{sat}=2N_cN_f$ and the maximum value of the Polyakov loop $\langle|L|\rangle $ which is a bit higher for $N_f=2$. The direct comparison of the Polyakov-loop expectation values in Fig.~\ref{vlg_poly} shows that the deconfinement crossover tends to start at somewhat smaller values of $\mu$ for $N_f=2$, and it gets quenched later, likewise. 

\begin{figure}
 \includegraphics[width=.9\linewidth]{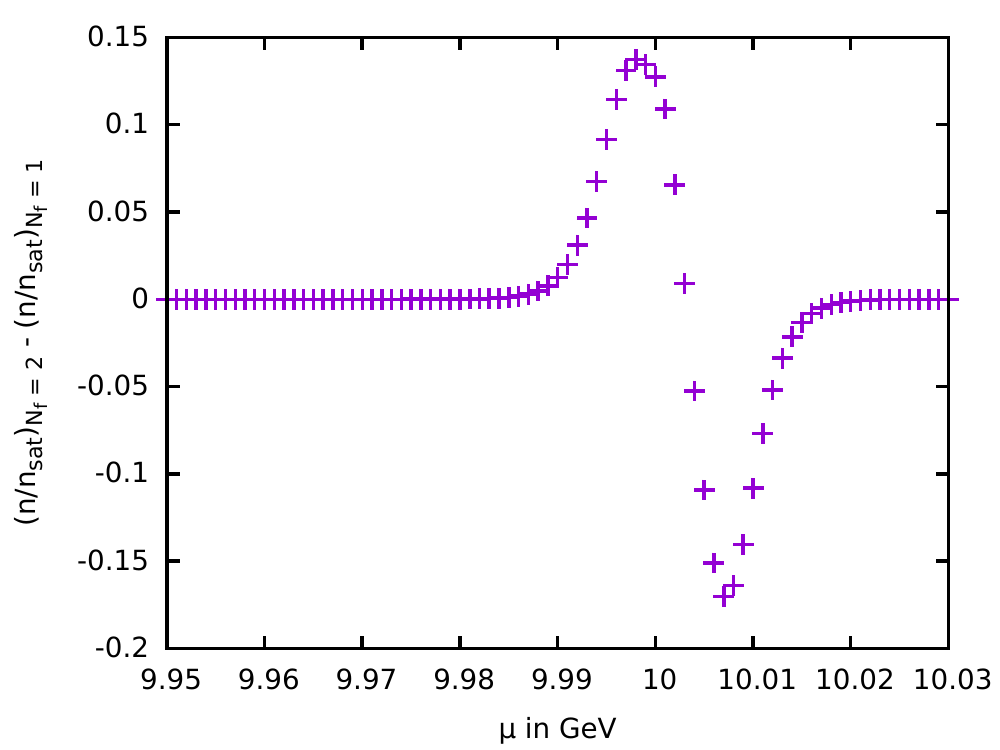}%
 \caption{Difference between the normalized quark number densities for $N_f=2$ and $N_f=1$ at $T=5$ MeV. \label{vlg_dens}}
 \end{figure}

The difference between the normalized quark-number densities ${n}/{n_\text{sat}}$ for $N_f=2$ and $N_f=1$ is shown in Fig.~\ref{vlg_dens}. We can see a deviation around $\mu=10$ GeV. This is in line with the observation that the deconfinement transition happens earlier for $N_f=2$ as well.

\begin{figure}
 \includegraphics[width=.9\linewidth]{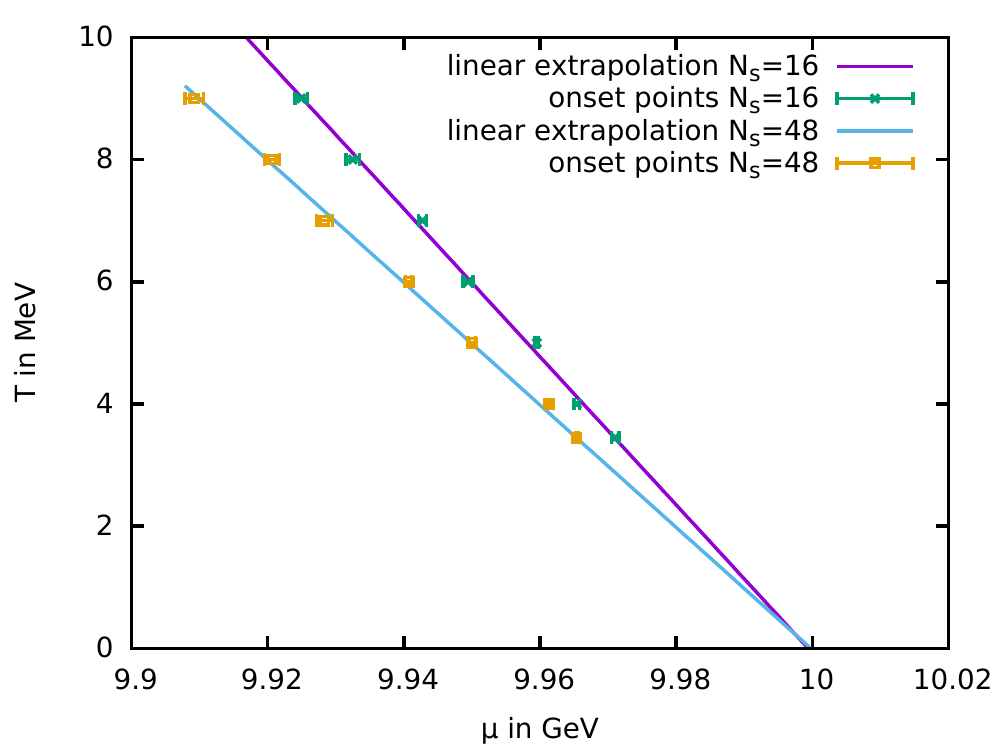}%
 \caption{Section of the $N_f=2$ phase diagram with a linear extrapolation of the diquark-density onset to $T=0$ at $\beta = 2.5$ or $a=0.081$~fm with $N_s=16$ and $N_s = 48$. \label{extrapol}}
\end{figure}

As described for $N_f=1$ above, we follow the same procedure with $N_f=2$ for various temperatures from $9$~MeV down to $3.454$~MeV. That is, for each temperature we determine the intersection point of the two exponential regimes in the quark density (for $N_f=2$ their $\mu$-values are consistently about $1-2$ MeV lower than those for $N_f=1$). Since the mass-parameter in the second exponential for $N_f=2$, see Eq.~(\ref{mfitNF2}), agrees well with the scalar diquark mass, $m_d=20$~GeV from Eq.~(\ref{mass}), we now take the intersection of the lines in logarithmic plots analogous to Fig.~\ref{slopes} as the onset of baryonic diquark matter and extrapolate the corresponding onset chemical potentials $\mu_c$ to $T=0$. The result for the $N_s=16$ lattice is shown in Fig.~\ref{extrapol}. Using a linear extrapolation as in the figure, which is consistent with a temperature independent $\tilde L \approx \langle L\rangle/2$, the result for the  $T=0$ diquark onset on the $N_s =16 $ lattice is $\mu_c = 9.9996(22)$~GeV and hence includes $m_d/2 = 10$~GeV within the error. Larger lattices lead to smaller values of $\tilde L$ and hence a smaller slope $\ln\tilde L$ in the linear extrapolation, but the extrapolated $\mu_c $ remains the same. For comparison, the same analysis was also done on a $N_s = 48$ lattice with the result that $\mu_c = 9.9998(9) $~GeV as also shown in Fig.~\ref{slopes}.

In order to test the scaling of this onset we have performed the same analysis also for 7 different lattice couplings $\beta $ between 2.4 and 2.5, corresponding to lattice spacings between $a= 0.1124$~fm and $0.0810$~fm with $\kappa$ values adjusted so that $m_d$ from Eq.~(\ref{mass}) remains fixed at 20~GeV as before. Again, for each $\beta$ we extract the corresponding intersection points of the two exponential regimes in the quark density at the same 7 temperatures between 9~MeV and 3.454~MeV. The extrapolated $N_f=2$ results for the zero-temperature diquark-density onsets from these intersection points are collected in Fig.~\ref{spacings}. Within the errors, these extrapolated values for $\mu_c$  basically all agree. Assuming that $\mu_c$ is thus independent of the lattice spacing in this parameter regime we simply use their average as our final overall estimate of
\begin{equation}
  \mu_c = 10.0001(3)
\end{equation}
from the data in Fig.~\ref{spacings} as indicated by the horizontal line with the gray error band. This overall estimate thus confirms that $\mu_c = m_d/2$ with rather high precision.

This agrees with the corresponding onset of isospin density at $m_\pi/2$ in the effective theory for heavy quarks in QCD \cite{Langelage},  and it shows that there is no ``Silver Blaze'' problem \cite{Cohen:2003kd} in the effective lattice theory for two-color QCD with heavy quarks either. Unlike the nuclear-matter transition at $3\mu_c = m_B - \epsilon$, with some evidence of a finite binding energy per nucleon $\epsilon$ in the effective theory for heavy quarks in QCD \cite{Langelage}, there is certainly no such evidence of a shift of the onset $\mu_c$ by a non-zero $\epsilon $ here. This is consistent with the generally expected difference between a first-order liquid-gas transition in QCD and Bose-Einstein condensation of diquarks in a second-order transition in two-color QCD. At the same time, however, the diquark densities obtained here are far from reflecting any sign of Bose-Einstein condensation. Quarks and diquarks are way too heavy to interpret the latter as deeply bound dimers. With our parameters from Eq.~(\ref{dbind}) the 20~GeV diquarks are only bound by about 2.8~MeV. If it wasn't for confinement, the transition temperature of the diquark-condensation phase by pair breaking should roughly be of the same order. Probably only because of the statistical confinement of the quarks in the first place, at the available temperatures above 4~MeV, all we can observe here therefore is an essentially free heavy-diquark gas behavior in the small window between $\mu = \mu_c(T) $ and the beginning deconfinement crossover followed by lattice saturation.  Unfortunately, the region where one might find a superfluid diquark-condensation phase thus is currently still beyond reach within the convergence region of the hopping series. Nevertheless, we can attempt to give a very rough first estimate of a region where such a diquark superfluid might be found, if we were able to further reduce the temperature, as follows: 

\begin{figure}
 \includegraphics[width=.9\linewidth]{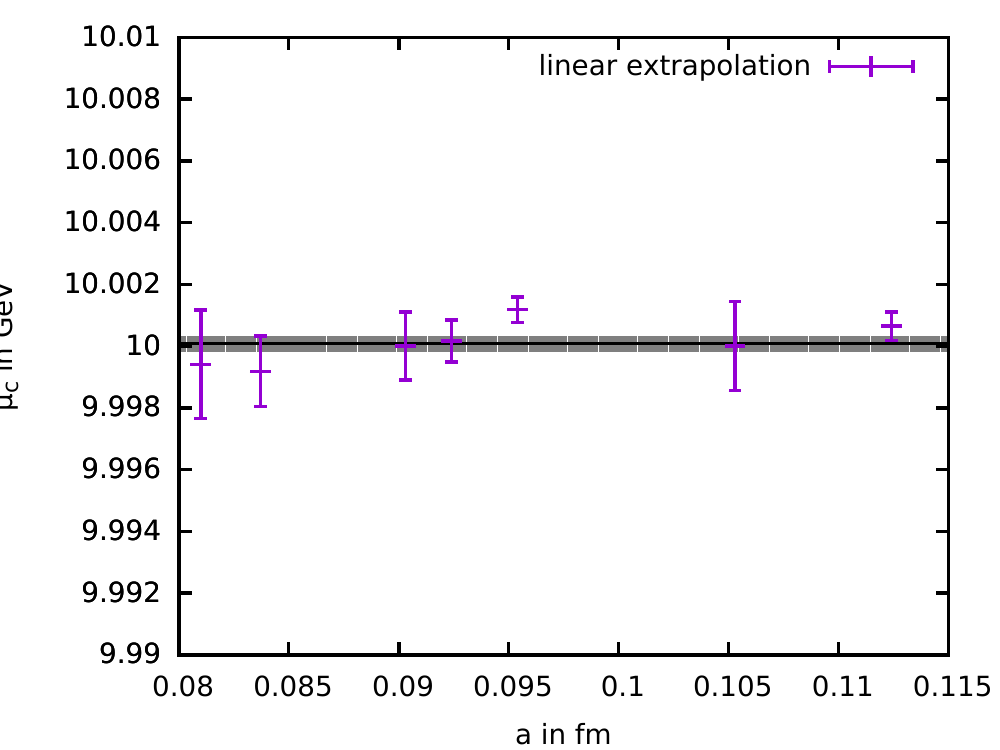}
 \caption{Critical chemical potentials $\mu_c$ for the diquark-density onset at $T=0$ from linear extrapolations for different lattice spacings between $a=0.0810$~fm and $0.1124$~fm corresponding to lattice couplings between $\beta = 2.5$ and $2.4$. \label{spacings}}
\end{figure}

\begin{figure}
 \includegraphics[width=.9\linewidth]{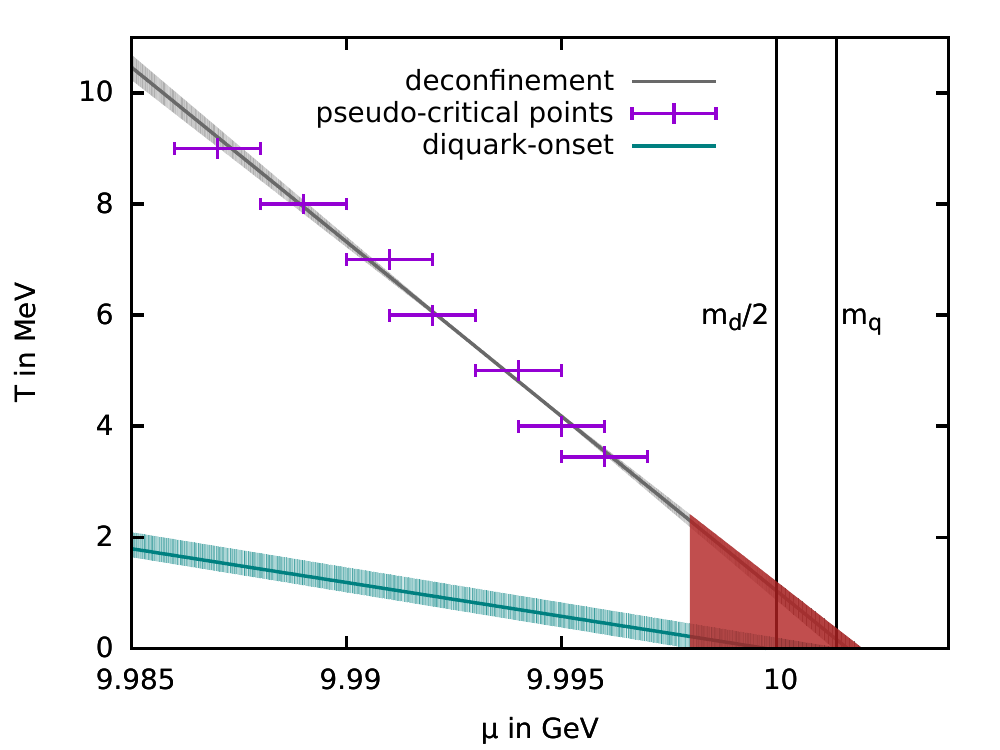}%
 \caption{Extrapolation of the deconfinement transition compared to the extrapolated diquark-density onset from Fig.~\ref{extrapol}. The region where one might hope to find a superfluid diquark-condensation phase is marked by the shaded red triangle. Half the diquark mass, $m_d/2 = 10$~GeV, and the quark mass $m_q = 10.0014$~GeV are indicated by vertical lines. \label{region}}
\end{figure}

Since the pseudo-critical chemical potentials for deconfinement at the available temperatures are also all below $m_d/2$, we compare their zero-temperature extrapolation to that of the diquark-density onset in Fig.~\ref{region}. The difference between the so extrapolated $\mu_c = m_d/2$ and $\mu_{pc}$ is small but significant. As seen in the figure, the deconfinement crossover then hits the $T=0$ axis of the phase diagram just above $m_q = 10.0014$~GeV. Therefore, a small window for a potential superfluid diquark-condensation phase at sufficiently low temperatures remains. The region where this might occur is indicated by the shaded red triangle in Fig.~\ref{region}. This region starts at a chemical potential slightly below $\mu=m_d/2=10$~GeV, i.e.~at the lower limit given by the extrapolation error of the diquark-density onset for the $\beta=2.5$ data used here. We use this lower limit instead of $m_d/2$ because there are also some truncation errors in our equations for the diquark mass $m_d$, Eq.~\eqref{mass}, and the effective fermion couplings $h$ and $h_2$ in Eqs.~\eqref{h1} and \eqref{h2}. The deconfinement transition temperature at this lower limit is then of roughly the same order as the diquark-binding energy and hence of the naive estimate of the transition temperature of a possible diquark-condensation phase.

 \section{Summary and outlook}
In this paper we have presented results for the baryonic diquark-matter onset in two-color QCD with heavy quarks. The results were obtained by numerical lattice simulations of a three-dimensional effective theory of Polyakov-loop variables analogous to that previously used for QCD \cite{Langelage}. The effective theory is thereby derived from the full theory by a combined strong-coupling and hopping expansion, in this case up to the combined order in $(u^n \kappa^m)$ with $n+m=4$. This effective theory was applied to the cold and dense regime of the phase diagram of two-color QCD with both $N_f=1$ and $N_f=2$. We have mapped the pseudo-critical line of the deconfinement crossover at small temperatures as indicated by the Polyakov-loop expectation value. With the scale set by the vacuum string tension as in QCD, we could reach temperatures ranging between $3.5$~MeV and $9$~MeV. At these low temperatures the Polyakov-loop expectation value $\langle L\rangle $ stays constant at its small but finite zero net-baryon density value for chemcial potentials up to the deconfinement crossover at around a pseudo-critical $\mu_{pc}$. Within this constant regime we could clearly identify a kink in the logarithm of the quark-number density where its finite-temperature behavior changes between that of a heavy-quark gas with imperfect statistical confinement in a finite volume and an essentially free gas of heavy diquarks with mass $m_d $ in a small window of chemical potentials $\mu$ between this diquark-density onset at $\mu_c \sim m_d/2 + T \ln\langle L\rangle$ (here with $\langle L\rangle \in [-1,1]$) and $\mu_{pc} $. For $\mu >\mu_{pc} $ the density describes quark matter that approaches its saturation value as the lattice gets filled with the maximum number of quarks per site  allowed by the Pauli principle, and as expected from previous finite-density studies \cite{Hands2006,Maas2012,Fromm:2012eb}.

This is all described quite well already by the leading-order analytic formula for the density together with a mean-field description for the Polyakov loop
provided we make the following two adjustments:

The constant Polyakov-loop expectation value $\langle L \rangle$ needed to describe the heavy-quark gas with imperfect statistical confinement for $\mu<\mu_c$, below the diquark onset, is only consistent with the very small but finite expectation value of the local Polyakov-loop variable $L_{\vec x}$ as obtained from its unquenched probablility distribution which is slightly distorted by the large but finite quark mass $m_q$. This local Polyakov-loop expectation value is about two orders of magnitude smaller in this constant regime than the expectation value of the modulus of its volume average $\langle |L| \rangle$ usually used as the order parameter.

In the heavy-diquark gas regime we have to replace $2m_q$ by $m_d$ as computed analytically in the combined strong-coupling and hopping expansion. Especially for $N_f=2$ the relevant mass $m_d$ here is that of a scalar diquark degenerate with the pion. With our parameters we only have $2 m_q - m_d = 2.8$~MeV as the binding energy of this scalar diquark but that makes a significant difference in the exponential increase of the diquark density for $\mu_c < \mu < \mu_{pc}$.

As soon as the Polyakov-loop starts to rise from its zero-density value, $\langle L_{\vec x} \rangle $ and $\langle |L|\rangle$ approach one another very rapidly.  For $\mu > \mu_{pc}$ we have $\langle L_{\vec x} \rangle = \langle |L|\rangle$ and their common $\mu$-dependent value in the leading-order density
describes the approach towards lattice saturation qualitatively quite well.

The leading-order formula for the density can also be calculated analytically without the mean-field prescription for the Polyakov loop, of course. Then, however, it has only a single exponential increase and there is no diquark onset just as there is no difference between $m_d$ and $2 m_q$. Furthermore, one can also calculate the Polyakov-loop variable $\tilde L$ in the mean-field formula from the exact leading order density for comparison. Its maximum at $\mu = m_q = -\ln(2\kappa)$ for $N_f=1$ then only results to be $\tilde L_\mathrm{max} = \sqrt{17/16} - 1 \approx 0.031 $, for example, which is more than an order of magnitude smaller than that of the measured Polyakov-loop expectation value. 

On the other hand, with the adjustments mentioned above the leading-order mean-field formula is flexible enough to describe the $\mathcal O(\kappa^4)$-data. The agreement between the measured quark-number density and the mean-field treatment of the Polyakov loop in the leading-order form might in fact be 
somewhat surprising, because the measured Polyakov-loop distribution is not at all sharply peaked around the mean-field value. It is a rather broad distribution suggesting non-negligible fluctuations around the mean value. A similar effect for an effective Polyakov-loop model with dynamical fermions in $SU(3)$  was found in \cite{Greensite2014}.

The relatively detailed understanding of the quark number density allows us to interpret the most interesting baryonic regime $\mu_c< \mu< \mu_{pc} $ as an essentially free heavy-diquark gas with some confidence. There is no direct evidence of those baryonic diquarks forming a Bose-Einstein condensate (BEC), however. The temperatures from $3.5$~MeV upwards are still too high for our 20~GeV diquarks bound by only about 2.8~MeV. Our relatively stable $T\to 0$ extrapolations of diquark-density onset and the deconfinement crossover nevertheless allowed to identify a small region between the two where one might hope to find signs of a BEC in the future.  

The fact that the $T=0$ extrapolation of the diquark-density onset $\mu_c$ is scale independent and from our overall average, covering various different lattice couplings between $\beta = 2.4$ and 2.5, agrees with $m_d/2$ within a relative accuracy of $2\cdot 10^{-5}$ at least is a strong indication that the zero-temperature diquark matter forming above $\mu_c = m_d/2$ in two-color QCD does not involve binding energy which would shift the onset to values smaller than the lightest baryon mass as in QCD. In the future one might think about including  a diquark source term and analyzing also a possible diquark condensate. In order to assess how much we can stretch the effective theory towards lighter masses and/or lower temperatures, which both increases the expansion parameter of the effective theory, it might be instructive to compare our results to full two-color QCD simulations with heavy quarks at finite $\mu$. Analogous work for $G_2$-QCD with fermionic baryons and evidence of a first-order liquid-gas transition of $G_2$-nuclear matter \cite{Wellegehausen:2015iea} is currently in progress.

\begin{acknowledgments}
We thank Georg Bergner, Owe Philipsen, Jochen Wambach and Bj\"orn Wellegehausen for helpful discussions. This work is supported by the Helmholtz International Center for FAIR  within the LOEWE program of the State of Hesse and the Helmholtz Graduate School HGS-HIRe. 
\end{acknowledgments}

\bibliography{Dichte}

\end{document}